\documentclass[journal,compsoc]{IEEEtran}
\usepackage{enumitem}
\usepackage{enumitem}
\usepackage{amsmath}
\usepackage[table]{xcolor}
\usepackage{graphicx}
\usepackage{multirow}
\usepackage{tabularx}
\usepackage{bigstrut}
\usepackage{booktabs} 
\usepackage{numprint}

\newcounter{magicrownumbers}
\newcommand\rownumber{\stepcounter{magicrownumbers}\arabic{magicrownumbers}}

\renewcommand{\theequation}{\arabic{equation}}

\newcommand{\tabref}[1]{Table~\ref{tab:#1}}

\newcommand{\figref}[1]{Figure~\ref{fig:#1}}

\makeatletter
\def\tagform@#1{\maketag@@@{(\ignorespaces\textit{Eq.\ #1}\unskip\@@italiccorr)}}
\renewcommand{\eqref}[1]{\textup{{\normalfont(\ref{#1}}\normalfont)}}

\makeatother

\ifCLASSOPTIONcompsoc
  \usepackage[nocompress]{cite}
\else
  \usepackage{cite}
\fi

\definecolor{darkred}{rgb}{0.5,0,0}    
\definecolor{darkgreen}{rgb}{0,0,0}   	
\definecolor{darkblue}{rgb}{0,0,0.5}
\definecolor{airforceblue}{rgb}{0.36, 0.54, 0.66}
\definecolor{arsenic}{rgb}{0.23, 0.27, 0.29}
\definecolor{cadetblue}{rgb}{0.37, 0.62, 0.63}

\usepackage[colorlinks,bookmarksnumbered,linkcolor=darkblue,citecolor=darkred,urlcolor=darkgreen]{hyperref}

\begin{document}
%
\title{An Analysis of Python's Topics, Trends, and Technologies Through Mining Stack Overflow Discussions}
%
%
%

\author{Hamed~Tahmooresi,
        Abbas~Heydarnoori, and
        Alireza~Aghamohammadi
        
\IEEEcompsocitemizethanks{\IEEEcompsocthanksitem H. Tahmooresi, A. Heydarnoori, and A. Aghamohammadi are with the Department
of Computer Engineering, Sharif University of Technology, Iran.\protect\\
E-mail: heydarnoori@sharif.edu
 \protect\\
E-mail: tahmooresi@ce.sharif.edu\protect\\
E-mail: aaghamohammadi@ce.sharif.edu
}}

%
%

\markboth{}
{}
%



\maketitle

\begin{abstract}
Python is a popular, widely used, and general-purpose programming language. 
In spite of its ever-growing community, researchers have not performed much analysis on Python's topics, trends, and technologies {\color{black} which provides insights for developers about Python community trends and main issues.}
In this article, we examine the main topics related to this language being discussed by developers on one of the most popular Q\&A websites, Stack Overflow, as well as temporal trends through mining \numprint{2461876} posts.
To be more useful for the software engineers, we study what Python provides as the alternative to popular technologies offered by common programming languages like Java.
Our results indicate that discussions about Python \textit{standard features}, \textit{web programming}, and \textit{scientific programming}
\footnote{Programming in areas such as mathematics, data science, statistics, machine learning, natural language processing (NLP), and so forth.}
are the most popular areas in the Python community. At the same time, areas related to the scientific programming are steadily receiving more attention from the Python developers.
\end{abstract}


\IEEEpeerreviewmaketitle

\section{Introduction}

\IEEEPARstart{P}{ython} is a widely used, high-level, general-purpose, interpreted, and dynamic programming language ~\cite{IEEESoftware_Redondo2015}.
According to Stack Overflow annual survey engaging more than \numprint{90000} developers in 2019, Python just edged out Java in overall ranking at the end of 2018, much like it edged out C\# in 2017, and PHP in 2016 alike. Stack Overflow called Python the \textit{fast-growing major programming language}~\cite{StackOverflow2019}.
In spite of the prevalence of Python, researchers have not adequately focused on analyzing the trends and technologies of this language in software developer communities, and Question and Answer (Q\&A) websites such as Stack Overflow {\color{black} which is the de facto website actively used by developers.
 Analyzing this invaluable resource of information can help developers to gain insight about Python's trends and main issues.}
Prior studies have leveraged a topic modeling approach called Latent Dirichlet Allocation (LDA)~\cite{LDA2003} to categorize discussions in specific areas such as \textit{web programming}~\cite{MSR_BajajPM2014}, \textit{mobile application development}~\cite{ESE_Rosen2016}, \textit{security}~\cite{jcst-YangLXWS216}, and \textit{blockchain}~\cite{wan2019discussed}. However, none of them has focused on Python. In this study, we shed light on the Python's main areas of discussion through mining \numprint{2461876} posts, {\color{black}from August 2008 to January 2019,} using LDA based topic modeling . We also analyze the temporal trends of the extracted topics to find out how the developers' interest is changing over the time. After explaining the topics and trends of Python, we investigate the technologies offered by Python, which are not properly revealed by the topic modeling. Because of the significant prevalence of Python, software engineers, mostly the experts of other programming languages, may be eager to know the technologies of their interest provided by Python as the alternative to investigate this language further.
That expert may search online resources and read a lot to  find an appropriate answer. However,
the aggregation and verification of the obtained knowledge may be on her own.
As a result, by leveraging the word embedding model~\cite{NIPS_MikolovSCCD2013}, we can find out for each technology provided for other programming languages what technologies are offered by Python which are practically employed by Python developers. Hence, we pave the way for better understanding Python's solutions. The main findings from our study indicate that Python developers mostly discuss Python \textit{standard features}, \textit{web programming}, and \textit{scientific programming}. However, the popularity of \textit{scientific programming} is growing faster.

\section{Study Setup} 
The primary goal of our study is to extract the areas that Python developers discuss on Stack Overflow. For this purpose, we introduce three research questions:
\begin{description}[style=unboxed,leftmargin=0cm]
\item[RQ1:]
What are the discussion areas in the Python community?
\item[RQ2:]
How is the interest of Python developers changing over the time?
\item[RQ3:] What technologies does Python provide?
\end{description}

\begin{figure*}[ht]
\centering
\includegraphics[width=0.9\linewidth]{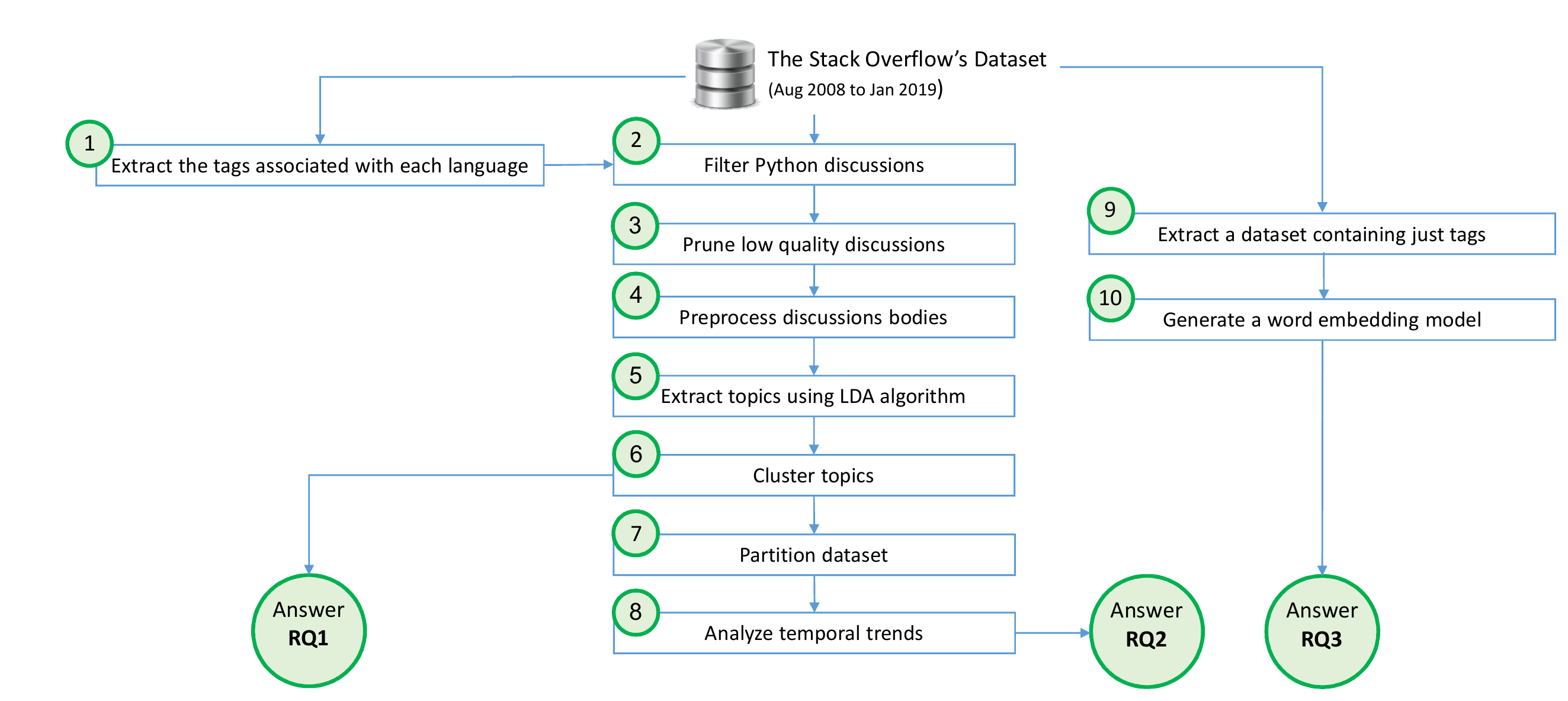}
\caption{Our high-level study methodology.}
\label{fig:methodology}
\end{figure*}

\figref{methodology} illustrates the high-level steps we take to answer our research questions.

\subsection{RQ1: What are the discussion areas in the Python community?}
To address \textbf{RQ1}, 
we first extracted the tags associated with each language (Step 1). Similar to prior studies, we identified discussions of each language according to Stack Overflow's tag mechanism~\cite{MSR_AllamanisS13, ESE_Barua2014}. In this website, the owner of a question may assign, up to five tags as keywords to each question which denotes the technological categories of that question~\cite{ESE_Barua2014}.
We preprocessed Python posts extracted (Step 2) as the input of our topic modeling. 

We pruned questions with the minus or zero score without any accepted answers in order to improve the overall quality of the topic modeling (Step 3). Due to the scoring mechanism, these questions have poor quality~\cite{MSR_BajajPM2014}.
This way, 9\% of all the Python posts were pruned from our dataset. Next, we cleansed a textual content of the extracted posts (Step 4) as follows:
\begin{enumerate}
\item All code snippets enclosed in the \texttt{<code>} tag were discarded since the source code would introduce noise in the results of the topic modeling~\cite{ICSE_Thomas2011}.
\item
All HTML tags were removed as well (e.g., \texttt{<a href="...">}, \texttt{<b>}, and so forth).
\item
We teokenized the corpus and removed common English-language stop words such as \texttt{a}, \texttt{the}, and \texttt{is} which do not help creating meaningful topics.
\item
All the tokens were stemmed using the Porter stemming algorithm~\cite{Program_Porter06}.
\end{enumerate}
Afterwards, we exploited the popular topic modeling technique, LDA, to investigate the developers' discussion areas (Step 5). LDA is an unsupervised model for performing statistical topic modeling that uses a \textit{bag of words} approach~\cite{LDA2003}. Recently, LDA has been employed in software engineering communities such as Stack Overflow to extract topics discussed in the crowd~\cite{MSR_BajajPM2014,ESE_Rosen2016}.
It takes the number of topics $K$ as an input. Larger values of $K$ will result in fine-grained topics and lower values will produce coarse-grained topics. Unfortunately, by choosing a small $K$ (e.g., 10 or 20), many discussion areas may remain hidden. As an example, an LDA model with 100 topics for Python discussions reveals topics such as \textit{game programming}, \textit{IoT}, and \textit{testing}. However, another model with 20 topics suppresses these areas. On the other hand, as the number of topics increases, their visualization and analysis become hard to understand. Besides, many duplicates emerge among topics, and thus a manual merge is required~\cite{ESE_Rosen2016}. Therefore, we first generated a LDA model with 100 topics and then asked two Python experts --- who are not the authors of this paper --- with more than seven years of experience to manually merge the LDA model.

The experts observed that topics extracted from Stack Overflow usually obtain known technologies among their top probable words (e.g., \textit{XML}, \textit{server}, \textit{web framework names}, and so forth). Therefore, if two topics share similar words having high occurrence probability, which match the name of known technologies, they may be good candidates for being merged. For example, if two topics have the word \textit{django}  --- a Python web framework --- as a highly probable word, we can anticipate that they are both about web programming subject. As a result, the experts considered Stack Overflow tags as a source of words resembling technologies to help merge topics more accurately. Having examined the Stack Overflow tags along with fine-grained topics, our experts decided to merge topics into 12 clusters (Step 6). Note that, the experts operated independent of each other. Then, they shared the results. In the case of difference, they discussed until all of the disagreements were resolved.   
\subsection{RQ2: How is the interest of Python developers changing over the time?}
To address \textbf{RQ2}, we partitioned the dataset (obtained in Step 2) into three-month time intervals (Step 7). Three-month intervals yielded enough discussions in each interval which make our analysis more reliable. Since the Python community is constantly growing, we observed that over the time, the frequency of each cluster is increasing as well. Thus, to better analyze the data, we borrowed the concept of \textit{impact} of a topic presented by Barua et al.~\cite{ESE_Barua2014} to obtain the portion of a topic $t_k$ in a time interval $intv$:
\begin{equation}\label{eq:impact}
    impact(t_k,intv) = \frac{1}{D(intv)}\sum_{d_j \in D(intv)}\theta(d_j,t_k)
\end{equation}
Where $D(intv)$ is the set of all posts in the time interval $intv$. $\theta(d_j,t_k)$ denotes the probability of a particular topic
$t_k$ occurring in the document $d_j$. In order for calculating the impact of each cluster in an interval, we simply sum up the impact of its topic.

Note that, a decrease in the impact of a topic does not mean a decline in discussions related to it. In fact, $impact(t_k,intv)$ resembles the portion of a topic in an interval to other topics as Equation \eqref{eq:impact} contains a fraction over $D(intv)$. Therefore, if other topics grow faster, the topic loses its portion. Now we can analyze the interest of developers on each topic cluster over the time (Step 8). We exploited the \textit{MK test} to find an increase or decrease in trends of the clusters. The \textit{MK test} is a non-parametric statistical method which assesses the existence of a monotonic increasing or decreasing trend (either linear or nonlinear) for a variable over the time~\cite{JSTOR_mann1945}.

\subsection{RQ3: What technologies does Python provide?}
We consider technology as software solutions, packages, libraries, and frameworks provided for developers in a programming language such as \textit{Django}~\cite{ESE_Rosen2016}. In this section, we focus on Python's technologies as alternatives to popular solutions provided by common programming language like Java. {\color{black} Hence, experts of the other programming languages can better get familiar with Python technologies which are considered alternatives to the technologies they use.} To this aim, we exploit the same idea suggested by Chen et al.~\cite{ESE_ChenXL19} that uses the crowd knowledge to recommend corresponding technologies in defferent programming languages. The proposed approach is based on \textit{word2vec}, a computationally efficient predictive model for learning word embedding from a raw text. What \textit{word2vec} produces is a vector for each word in the corpus~\cite{NIPS_MikolovSCCD2013}. One of the interesting features of the model is to organize technologies and comprehend the implicit relationships between them via unsupervised learning~\cite{NIPS_MikolovSCCD2013}. For instance, suppose $V_{ph}$, $V_{l}$, $V_{py}$, and $V_d$ be the vector representations of the word PHP, Laravel, Python, and Django respectively. The results of the calculation $V_{l} - V_{ph} + V_{py}$ is closer to $V_d$ than any other word vectors. That is to say, not only do Django and Laravel have clusters near each other, but they each have similar distances in a vector space to the programming language whose web frameworks they are. Therefore, without even specifying any context (being the web framework of a programming language in this example), the algorithm can extract the latent relationship between languages.
We use the Stack Overflow's tag mechanism as developers use it to elaborate upon their libraries, frameworks, and technical concepts to which their questions are directly relative, in order to be easily found by respondents~\cite{ESEM_ChenX16}. Leveraging the tag mechanism to describe technologies in the software development community has been recently performed by Barua et al.~\cite{ESE_Barua2014}. Thus, we created a new dataset (Step 9) {\color{black}by using tags of each question}. For example, if a question is assigned the tags \textit{$<$java, swing, jframe$>$}, {\color{black} we simply consider "java swing jframe" as an item of our dataset. Finally, we trained our word embedding model using our dataset} (Step 10).
\section{Results}
Herein, we present the results and findings of our study.
\subsection{Areas Discussed by Python Developers}
\begin{figure}[ht]
    \centering
    \includegraphics[width=0.9\linewidth]{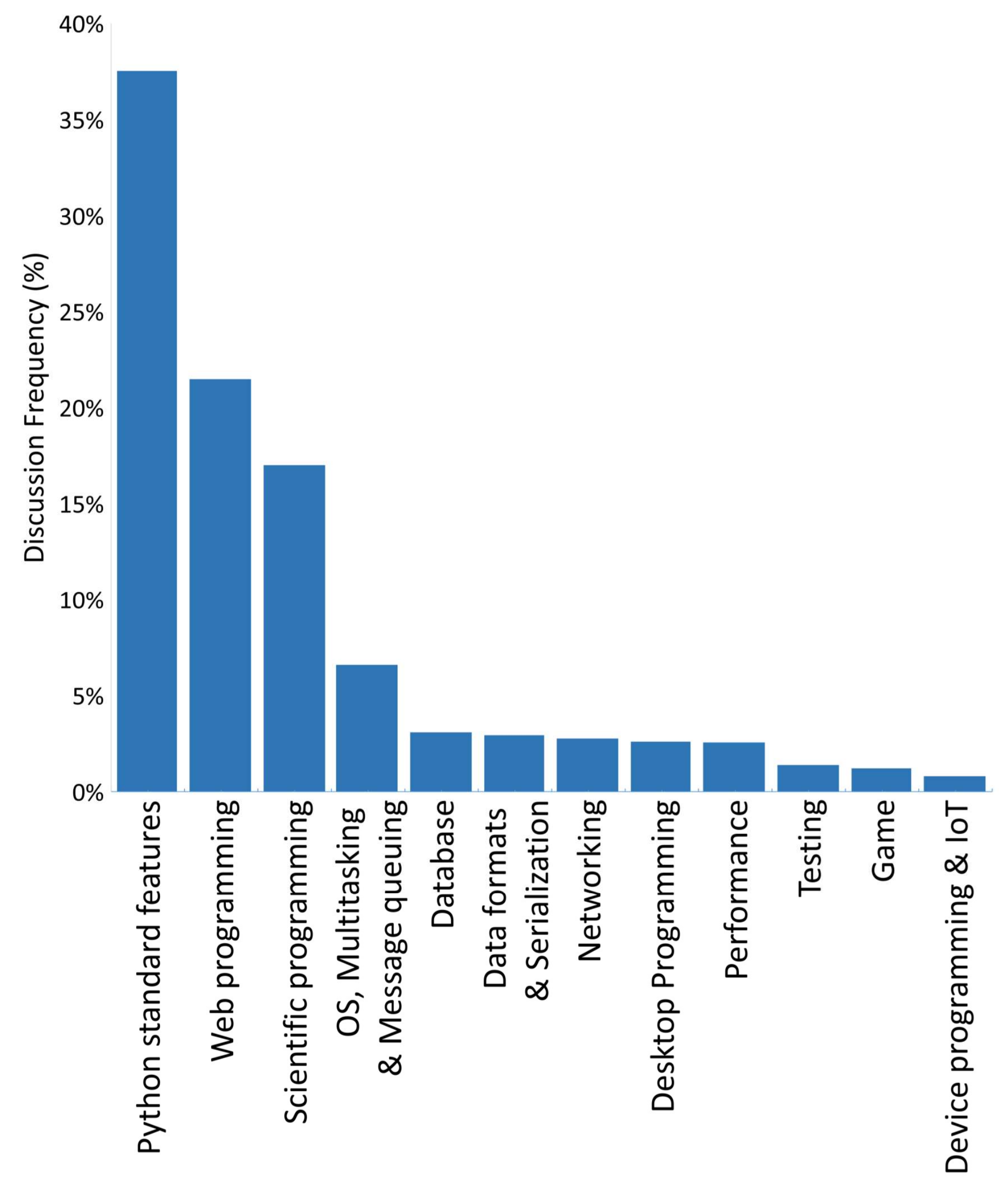}
    \caption{Categories of discussions related in Python.}
    \label{fig:python-clusters}
\end{figure}

As mentioned before, we used LDA to extract fine-grained areas discussed. Next, two experts grouped the topics into 12 clusters. Extracted Python clusters along with their portions are demonstrated in \figref{python-clusters}. The most frequent area belongs to \textit{standard features} and problems related to the Python language itself. This may be due to the fact that the Python community contains a significant number of amateur developers who have started Python without enough programming knowledge. The second and third most frequent discussion areas belong to the \textit{web programming} and \textit{scientific programming}. Furthermore, \textit{OS}, \textit{multitasking}, \textit{message queuing issued}, \textit{data format} (e.g., JSON, XML, CSV, etc.) and \textit{serialization} are also widespread among developers. Interestingly, Python developers are attracted to \textit{game programming} and issues about \textit{programming on devices} such as Rasberri Pi and Arduino which are popular in the IoT community.

\begin{figure}[ht]
    \centering
    \includegraphics[width=0.9\linewidth]{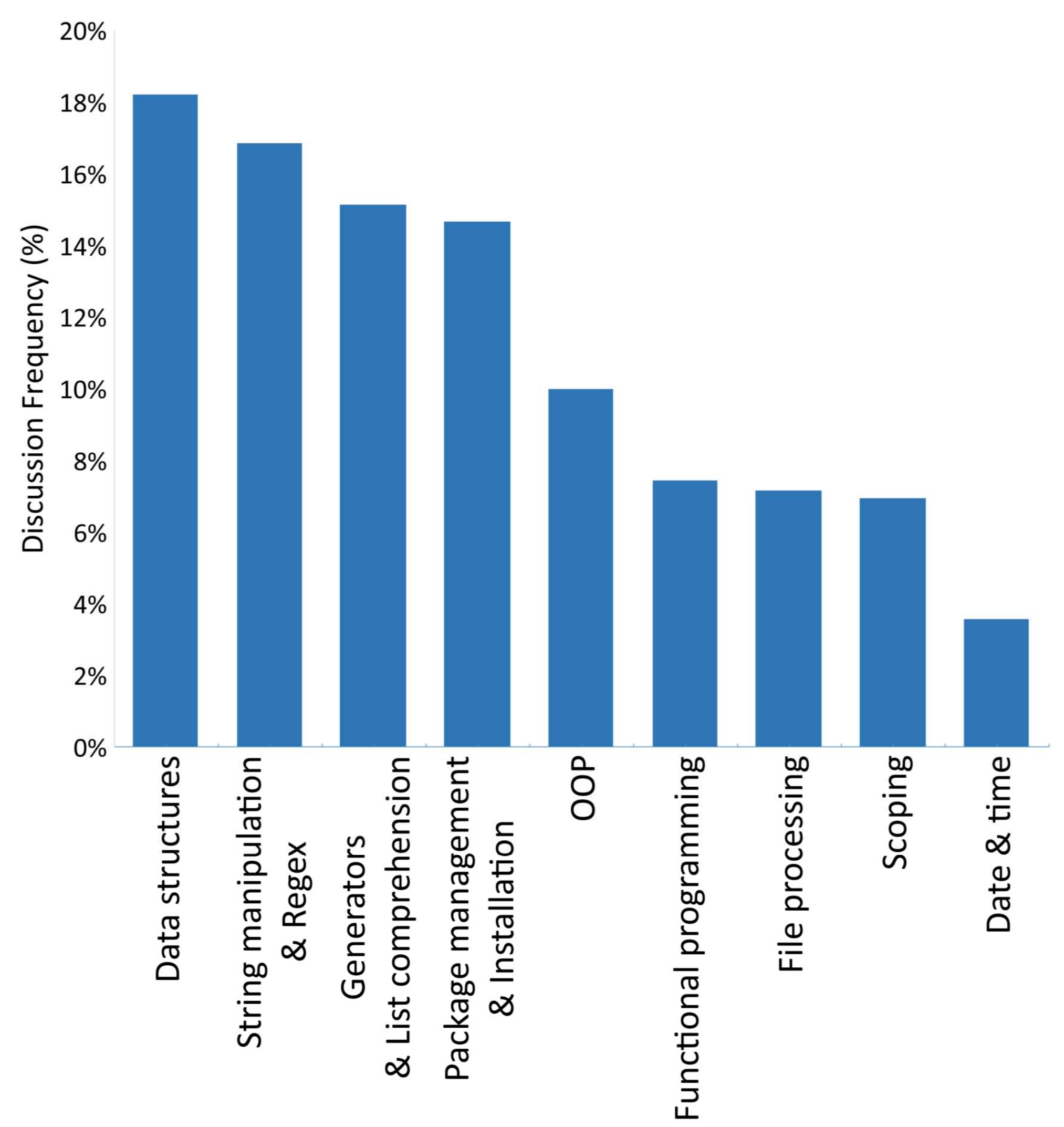}
    \caption{Topics inside the Python standard features cluster.}
    \label{fig:python-language-topics}
\end{figure}

Since we grouped topics into clusters, we can now simply drill a step down to the finer-grained areas. For example, \figref{python-language-topics} illustrates topics of the Python \textit{standard features} (The first bar in \figref{python-clusters}).
According to \figref{python-language-topics}, \textit{data structures}, working with \textit{strings}, \textit{list comprehension} and \textit{generators}, \textit{package management}, and \textit{installation} are the most discussed topics.
\textit{Object-oriented programming} (OOP) is the next popular area.

\begin{description}[style=unboxed,leftmargin=0cm]
\item[Study Findings.] Issues about Python \textit{standard features} are the most frequent area among Python programmers. After that, \textit{web} and \textit{scientific programming} are more popular than other areas.
\end{description}

\subsection{Temporal Trends}
\begin{figure}[ht]
    \centering
    \includegraphics[width=0.9\linewidth]{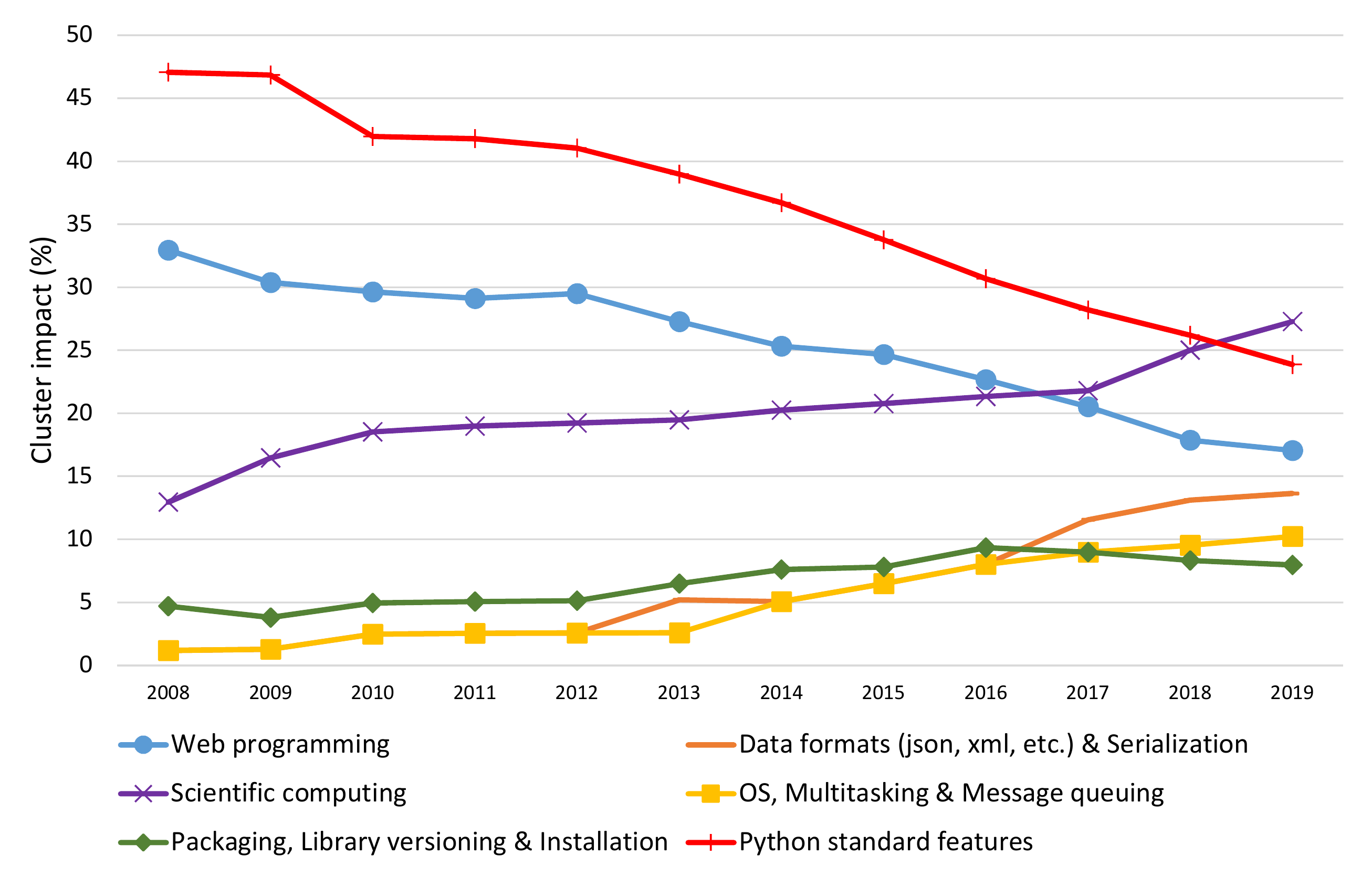}
    \caption{Temporal trends related to Python topic clusters.}
    \label{fig:temporal-trends}
\end{figure}

As for temporal trend analysis, we plotted only clusters having a significant trend (increasing or decreasing, according to \textit{MK test}) to obtain more beneficial results. As shown in \figref{temporal-trends}, discussions about Python standard features and \textit{web programming} are losing share among other areas. As a possible explanation, these two areas have been discussed from the very beginning of Stack Overflow, at the time when other areas were obscured. However, as other areas arise, the portions of the questions related to \textit{web programming} and Python \textit{standard features} decline over the time.
Besides, questions related to these two areas become saturated and most of the challenges are discussed which may amplify this decline.

\begin{description}[style=unboxed,leftmargin=0cm]
\item[Study Findings.]
\textit{Scientific programming} is increasing rapidly, exceeding \textit{web programming}. Challenges related to \textit{packaging}, \textit{library versioning}, and \textit{installation} have remained stable since 2015.
Another finding is that subject related to \textit{OS}, \textit{multitasking}, and \textit{message queuing} is rising as well as \textit{data formats} and \textit{serialization}, especially from the start of 2015.
\end{description}

\subsection{Python Technologies}
Leveraging Chen et al.'s work~\cite{ESE_ChenXL19}, we can now extract Python's technologies and organize them according to their correspondence to solutions offered by other programming languages. \tabref{word2vec-alternatives} shows a sample of the Python technologies recommended by the approach. {\color{black} However, developers can use our trained model (which is based on the approach offered by Chen et al.'s work~\cite{ESE_ChenXL19}) to further investigate Python's technologies.\footnote{\href{https://goo.gl/8Ne2cK}{https://goo.gl/8Ne2cK}}} As a simple example, given \textit{Visual Studio} --- the C\# IDE --- as an input, the approach suggests \textit{Pycharm} a popular Python IDE as a solution to the needs Visual Studio satisfies in the C\# community. As a more advanced example, the model recommends \textit{$<$Virtualenv, pip, requirements.txt$>$} as an alternative to \textit{Maven} (in Java). In reality, Python developers list their library dependencies in a file so-called \textit{requirements.txt} (analogous to \textit{pom.xml} in \textit{Maven}). Furthermore, by using \textit{pip}, all dependencies can be downloaded from the \textit{pip} online repositories (analogous to \textit{Maven} repositories) to a directory located in the host machine. However, there come two main differences between the Python's and Java's way of building applications:
\begin{enumerate}
\item Unlike \textit{Maven},  \textit{pip} does not keep multiple versions of the same package.
\item
All applications in the same environment consider a single directory as a source code of external packages. While in Java, each application has its \textit{lib} directory in which maintains a copy of libraries that depends on Jar files.
\end{enumerate}
As a result, there is no way to have multiple applications depending on different versions of the same package in a single machine. Here comes a popular tool, \textit{Virtualenv},  to create multiple virtual environments on the same machine. Each has its own Python interpreter, site-package, \textit{pip} tool, and so forth. 

Note that, experts of the other programming languages may use the Google search engine to find their correspondences offered by Python. For example, if we search "Java Maven alternative in Python" the first recommended page would be a Stack Overflow question that its accepted answer only mentions \textit{distutils} and \textit{setuptools}. In fact, the answer only covers packaging tools in Python and skips \textit{dependency management}. Furthermore, \textit{Pybuilder} is recommended in topic links, which is not popular as at the time of writing this paper. The frequency of its corresponding tag in the Stack Overflow is just 43.
However, using the word embedding approach will extract technologies which are real alternatives and practically used according to millions of discussions in the crowd~\cite{ESE_ChenXL19}.

\begin{table*}[t]
\caption{The catalog of Python's solutions including their correspondence to technologies in other languages. }
{\small
\begin{tabular}{ 	p{\dimexpr 0.05\linewidth-2\tabcolsep}	
					p{\dimexpr 0.15\linewidth-2\tabcolsep}	
					p{\dimexpr 0.40\linewidth-2\tabcolsep}	
					p{\dimexpr 0.40\linewidth-2\tabcolsep}	
				}
 & Technology & Context & Python's alternatives \bigstrut \\
\hline
\multicolumn{4}{c}{Java} \\
\hline
\hline
\rownumber & Hibernate & Object-relational mapping (ORM) & SQLAlchemy, Django-models \\
\hline
\rownumber & Maven & Build tool & Virtualenv, pip, requirements.txt \\
\hline
\rownumber & Swing & Desktop application development & PyQt, WXPython, TKinter \\
\hline
\rownumber & Stanford-NLP & Natural language processing & NLTK, Gensim \\
\hline
\rownumber & Jackson & JSON library \& object serializer & Pickle, SimpleJson, Django-serializer \\
\hline
\rownumber & Spring & Dependency injection, Module integration, Web development & Django, uwsgi, tornado, celery \\
\hline
\rownumber & Spring-MVC & Web programming & Django, Flask \\
\hline
\rownumber & JAR & Packaging & PyInstaller, Py2exe, Egg \\
\hline
\rownumber & Jackson & JSON library and object serializer & Pickle, simpleJson, Django-serializer \\
\hline
\rownumber & Eclipse & Integrated development environment (IDE) & Pycharm \\
\hline
\rownumber & Tomcat & Web server & Uwsgi, gunicorn, tornado, Nginx \\
\hline
\multicolumn{4}{c}{PHP} \\
\hline
\hline
\rownumber & Smarty & Template engine & Django-templates, Jinja2, Cheetah \\
\hline
\rownumber & PHPMailer & Mail sending & Sendmail, SMTPLib \\
\hline
\rownumber & Laravel & Web programming & Django, Flask \\
\hline
\rownumber & WordPress & Content management system & Django-CMS, Mezzanine \\
\hline
\rownumber & PDO & Data-access abstraction layer & MySQL-python, pymssql, psycopg2\\
\hline
\rownumber & cURL & transferring data using various protocols such as HTTP, FTP, and so forth.
& Python-requests\\
\hline
\multicolumn{4}{c}{C\#} \bigstrut \\
\hline
\hline
\rownumber & NUnit & Unit testing framework & Nose, Python-unittest \\
\hline
\rownumber & Visual Studio & Integrated Development Environment & Pycharm\\
\hline
\rownumber & Entity-framework & Object Relational Mapping (ORM) & Django-queryset, sqlachemy \\
\hline
\rownumber & ASP.NET-MVC & Web application development & Django, Flask\\
\hline
\rownumber & IIS & Web server & Uwsgi, gunicorn, tonado, Nginx\\
\hline
\rownumber & WFP & Rendering user interfaces in Windows-based applications & PyQt, WXPython, TKinter\\
\hline
\rownumber & Unity3d & Game engine & Pygame\\
\hline
\multicolumn{4}{c}{C \& C++} \bigstrut \\
\hline
\hline
\rownumber & Qt & Desktop application development & PyQt, WXPython, TKinter  \\
\hline
\rownumber & OpenCV & Image processing & Scikit, Pillow  \\
\hline
\multicolumn{4}{c}{Ruby} \bigstrut \\
\hline
\hline
\rownumber & rubygems & Package management & pip, Anaconda, virtualenv \\
\hline
\rownumber & Activerecord & Object-relational mapping (ORM) &  Django-queryset, sqlachemy \\
\hline
\rownumber & Cucumber & Test \& Behavior driven development & Lettuce, Robotframework \\
\hline
\rownumber & Devise & Authentication framework & Django-allauth, Django-authentication, Django-registration \\
\hline
\multicolumn{4}{c}{Javascript} \bigstrut \\
\hline
\hline
\rownumber & NPM & Package management & Pip, Anaconda, virtualenv\\
\hline
\rownumber & Socket.io & Realtime web application development & Tornado, gevent\\
\hline
\rownumber & Mongoose & MongoDB ODM (Object Document Mapper) & Pymongo\\
\hline
\rownumber & Sequelize & Object Relational Mapping (ORM) & SQLAchemy, Django-models\\
\end{tabular}
}
\label{tab:word2vec-alternatives}
\end{table*}

\section{Related Work}
Several studies have been performed using Stack Overflow for a wide variety of purposes.
Researchers have employed LDA to categorize discussions in specific areas such as \textit{web programming}~\cite{MSR_BajajPM2014}, \textit{mobile application development}~\cite{ESE_Rosen2016}, \textit{security}~\cite{jcst-YangLXWS216}, and \textit{blockchain}~\cite{wan2019discussed}.
For example, 
 Barua et al. used LDA to extract main topics discussed on Stack Overflow~\cite{ESE_Barua2014}. They also investigated how developer's interest would change over the time using the impact of a topic. Finally, to provide a more focused view, they extracted the change of interest in specific technologies over the time.
Rosen et al. concentrated on mobile developers on Stack Overflow which exploited LDA to extract main topics and most difficult issues of mobile developers~\cite{ESE_Rosen2016}.
However, none of the prior studies focused on Python despite of its ever-growing popularity.
\section{Conclusion}
In this article, we investigated topics and technologies of the Python programming language. We employed trend analysis and used a clustering approach for improving the understandability of a large number of topics. Furthermore, we used an approach based on the \textit{word2vec} model to recommend Python's solutions corresponding to technologies of other programming languages.

Our results indicate that \textit{standard features} provided by Python, \textit{web development}, and \textit{scientific programming} are the most popular areas among Python developers on Stack Overflow. However, \textit{scientific programming} is gaining more popularity. Ultimately, using the \textit{word2vec} model we extracted some of the Python's technologies as alternatives to technologies offered by other programming languages.


%

\ifCLASSOPTIONcaptionsoff
  \newpage
\fi



\bibliographystyle{IEEEtran}
\bibliography{IEEEabrv,ms}

\begin{thebibliography}{10}
\providecommand{\url}[1]{#1}
\csname url@samestyle\endcsname
\providecommand{\newblock}{\relax}
\providecommand{\bibinfo}[2]{#2}
\providecommand{\BIBentrySTDinterwordspacing}{\spaceskip=0pt\relax}
\providecommand{\BIBentryALTinterwordstretchfactor}{4}
\providecommand{\BIBentryALTinterwordspacing}{\spaceskip=\fontdimen2\font plus
\BIBentryALTinterwordstretchfactor\fontdimen3\font minus
  \fontdimen4\font\relax}
\providecommand{\BIBforeignlanguage}[2]{{%
\expandafter\ifx\csname l@#1\endcsname\relax
\typeout{** WARNING: IEEEtran.bst: No hyphenation pattern has been}%
\typeout{** loaded for the language `#1'. Using the pattern for}%
\typeout{** the default language instead.}%
\else
\language=\csname l@#1\endcsname
\fi
#2}}
\providecommand{\BIBdecl}{\relax}
\BIBdecl

\bibitem{IEEESoftware_Redondo2015}
J.~M. Redondo and F.~Ortin, ``A comprehensive evaluation of common python
  implementations,'' \emph{{IEEE} Software}, vol.~32, no.~4, pp. 76--84, 2015.

\bibitem{StackOverflow2019}
``{Developer Survey Results},''
  \url{https://insights.stackoverflow.com/survey/2019}, online; accessed 3 July
  2019.

\bibitem{LDA2003}
D.~M. Blei, A.~Y. Ng, and M.~I. Jordan, ``Latent dirichlet allocation,''
  \emph{Journal of machine Learning research,}, vol.~3, no. Jan, pp. 993--1022,
  2003.

\bibitem{MSR_BajajPM2014}
K.~Bajaj, K.~Pattabiraman, and A.~Mesbah, ``Mining questions asked by web
  developers,'' in \emph{Proceedings of the 11th Working Conference on Mining
  Software Repositories, {MSR}}.\hskip 1em plus 0.5em minus 0.4em\relax {ACM},
  2014, pp. 112--121.

\bibitem{ESE_Rosen2016}
C.~Rosen and E.~Shihab, ``What are mobile developers asking about? {A} large
  scale study using stack overflow,'' \emph{Empirical Software Engineering,},
  vol.~21, no.~3, pp. 1192--1223, 2016.

\bibitem{jcst-YangLXWS216}
X.~Yang, D.~Lo, X.~Xia, Z.~Wan, and J.~Sun, ``What security questions do
  developers ask? {A} large-scale study of stack overflow posts,'' \emph{J.
  Comput. Sci. Technol.}, vol.~31, no.~5, pp. 910--924, 2016.

\bibitem{wan2019discussed}
Z.~Wan, X.~Xia, and A.~E. Hassan, ``What is discussed about blockchain? a case
  study on the use of balanced lda and the reference architecture of a domain
  to capture online discussions about blockchain platforms across the stack
  exchange communities,'' \emph{IEEE Transactions on Software Engineering},
  2019.

\bibitem{NIPS_MikolovSCCD2013}
T.~Mikolov, I.~Sutskever, K.~Chen, G.~S. Corrado, and J.~Dean, ``Distributed
  representations of words and phrases and their compositionality,'' in
  \emph{Proceedings of the 27th Annual Conference on Neural Information
  Processing Systems}, 2013, pp. 3111--3119.

\bibitem{MSR_AllamanisS13}
M.~Allamanis and C.~A. Sutton, ``Why, when, and what: analyzing stack overflow
  questions by topic, type, and code,'' in \emph{Proceedings of the 10th Annual
  Conference on Mining Software Repositories}, 2013, pp. 53--56.

\bibitem{ESE_Barua2014}
A.~Barua, S.~W. Thomas, and A.~E. Hassan, ``What are developers talking about?
  an analysis of topics and trends in stack overflow,'' \emph{Empirical
  Software Engineering,}, vol.~19, no.~3, pp. 619--654, 2014.

\bibitem{ICSE_Thomas2011}
S.~W. Thomas, ``Mining software repositories using topic models,'' in
  \emph{Proceedings of the 33rd International Conference on Software
  Engineering, {ICSE}}.\hskip 1em plus 0.5em minus 0.4em\relax {ACM}, 2011, pp.
  1138--1139.

\bibitem{Program_Porter06}
M.~F. Porter, ``An algorithm for suffix stripping,'' \emph{Program}, vol.~40,
  no.~3, pp. 211--218, 2006.

\bibitem{JSTOR_mann1945}
H.~B. Mann, ``Nonparametric tests against trend,'' \emph{Econometrica: Journal
  of the Econometric Society}, pp. 245--259, 1945.

\bibitem{ESE_ChenXL19}
C.~Chen, Z.~Xing, and Y.~Liu, ``What's spain's paris? mining analogical
  libraries from q{\&}a discussions,'' \emph{Empirical Software Engineering},
  vol.~24, no.~3, pp. 1155--1194, 2019.

\bibitem{ESEM_ChenX16}
C.~Chen and Z.~Xing, ``Mining technology landscape from stack overflow,'' in
  \emph{Proceedings of the 10th International Symposium on Empirical Software
  Engineering and Measurement}, 2016, pp. 14:1--14:10.

\end{thebibliography}
%

%

\begin{IEEEbiography}[{\includegraphics[width=1.0in,height=1.25in,clip]{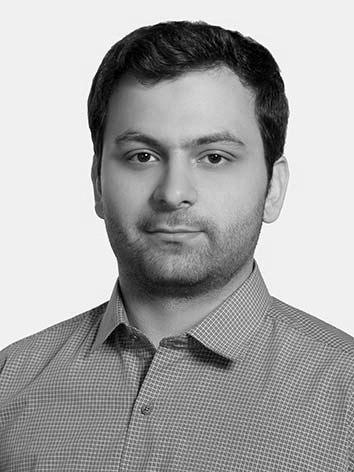}}]{Hamed Tahmooresi} is a PhD student at the Sharif University of Technology. His research interests include software engineering, software architecture and design, and mining software repositories. Contact him at \href{tahmooresi@ce.sharif.edu}{\nolinkurl{tahmooresi@ce.sharif.edu}}.
\end{IEEEbiography}

\begin{IEEEbiography}[{\includegraphics[width=1.0in,height=1.25in,clip]{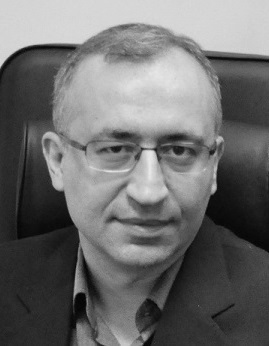}}]{Abbas Heydarnoori}
is an assistant professor in the Department of Computer Engineering at the Sharif University of Technology. Before, he was a post-doctoral fellow at the University of Lugano, Switzerland, working with Prof. Walter Binder. Abbas did his PhD in the School of Computer Science at the University of Waterloo, Canada, under the supervision of Prof. Krzysztof Czarnecki. His research interests focus on software evolution and maintenance, mining software repositories, and recommendation systems in software engineering. Contact him at
\href{mailto:heydarnoori@sharif.edu}{\nolinkurl{heydarnoori@sharif.edu}}.
\end{IEEEbiography}

\begin{IEEEbiography}[{\includegraphics[width=1.0in,height=1.25in,clip]{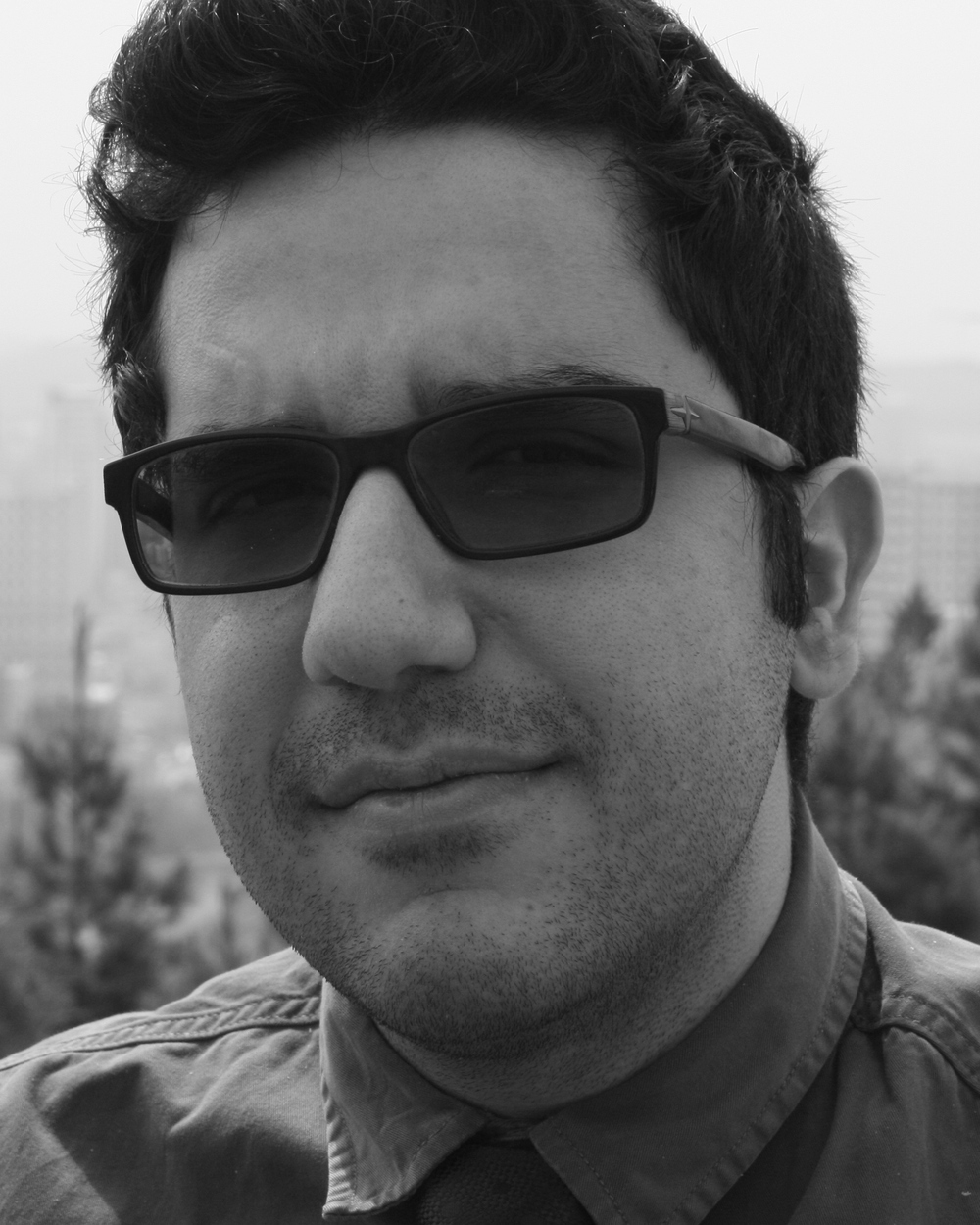}}]{Alireza Aghamohammadi}
is a Ph.D. student at Sharif University of Technology (SUT). He works on a wide range of recommendation systems that exploit state-of-the-art machine learning techniques. He is also a software developer for more than seven years. Contact him at \href{mailto:aaghamohammadi@ce.sharif.edu}{\nolinkurl{aaghamohammadi@ce.sharif.edu}}.
\end{IEEEbiography}




\end{document}